\documentclass[12pt]{article}
\usepackage{epsfig}
\usepackage{psfrag}
\usepackage{latexsym}
\usepackage[usenames,dvipsnames]{color}
\usepackage[hang,small,bf]{caption}
\usepackage[bookmarks,colorlinks]{hyperref}
\hypersetup{%
 pdftitle={Neutrino Masses with Inverse Hierarchy from Broken L_e-L_mu-L_tau: a Reappraisal},
pdfauthor={Guido Altarelli, Roberto Franceschini},
pdfkeywords={neutrino mass, inverse hierarchy},
bookmarksnumbered,
bookmarksopen=true,
pdfstartview={FitB},
urlcolor=Black,
linkcolor=Blue,
citecolor=Brown,
} %

\def\beq{\begin{equation}}
\def\eeq{\end{equation}}
\def\bea{\begin{eqnarray}}
\def\eea{\end{eqnarray}}
\def\bet{\begin{tabular}}
\def\eet{\end{tabular}}
\def\bes{\begin{subequations}\bea}
\def\ees{\eea\end{subequations}}

\newcommand{\hepph}[1]{\href{http://xxx.lanl.gov/abs/hep-ph/#1}{\tt hep-ph/#1}}

\def\be{\begin{equation}}
\def\ee{\end{equation}}
\def\bc{\begin{center}}
\def\ec{\end{center}}
\def\bea{\begin{eqnarray}}
\def\eea{\end{eqnarray}}

\catcode`@=11
\def\marginnote#1{}
\newcount\hour
\newcount\minute
\newtoks\amorpm
\hour=\time\divide\hour by60
\minute=\time{\multiply\hour by60 \global\advance\minute by-\hour}
\edef\standardtime{{\ifnum\hour<12 \global\amorpm={am}%
        \else\global\amorpm={pm}\advance\hour by-12 \fi
        \ifnum\hour=0 \hour=12 \fi
        \number\hour:\ifnum\minute<10 0\fi\number\minute\the\amorpm}}
\edef\militarytime{\number\hour:\ifnum\minute<10 0\fi\number\minute}
\def\draftlabel#1{{\@bsphack\if@filesw {\let\thepage\relax
   \xdef\@gtempa{\write\@auxout{\string
      \newlabel{#1}{{\@currentlabel}{\thepage}}}}}\@gtempa
   \if@nobreak \ifvmode\nobreak\fi\fi\fi\@esphack}
        \gdef\@eqnlabel{#1}}
\def\@eqnlabel{}
\def\@vacuum{}
\def\draftmarginnote#1{\marginpar{\raggedright\scriptsize\tt#1}}
\def\draft{\oddsidemargin 0.0truein
        \def\@oddfoot{\sl preliminary draft \hfil
        \rm\thepage\hfil\sl\today\quad\militarytime}
        \let\@evenfoot\@oddfoot \overfullrule 3pt
        \let\label=\draftlabel
        \let\marginnote=\draftmarginnote
   \def\@eqnnum{(\theequation)\rlap{\kern\marginparsep\tt\@eqnlabel}%
\global\let\@eqnlabel\@vacuum}  }
\catcode`@=12

\begin{document}
\begin{titlepage}
\vspace*{-1cm}
\phantom{hep-ph/***} 
\hfill{CERN-PH-TH/2005-250}
\vskip 0.1cm
\hfill{    RM3-TH/05-13}
\vskip 0.5cm
\begin{center}
{\Large\bf Neutrino Masses with Inverse Hierarchy from\\ Broken $L_e-L_\mu-L_\tau$: a Reappraisal}
\end{center}
\vskip 0.2  cm
\vskip 0.5  cm
\begin{center}
{\large Guido Altarelli}~\footnote{e-mail address: \href{mailto:guido.altarelli@cern.ch}{guido.altarelli@cern.ch}}
\\
\vskip .1cm
CERN, Department of Physics, Theory Division
\\ 
CH-1211 Geneva 23, Switzerland
\\
\vskip .1cm
and
\\
Dipartimento di Fisica `E.~Amaldi', Universit\`a di Roma Tre
\\ 
INFN, Sezione di Roma Tre, I-00146 Rome, Italy
\\
\vskip .2cm
{\large Roberto Franceschini}~\footnote{e-mail address: \href{mailto:roberto.franceschini@sns.it}{roberto.franceschini@sns.it}}
\\
\vskip .1cm
Dipartimento di Fisica `E.~Amaldi', Universit\`a di Roma Tre I-00146 Rome, Italy
\\ and \\
Scuola Normale Superiore - Piazza dei Cavalieri 7 -
I-56100 Pisa, Italy
\end{center}
\vskip 0.7cm
\begin{abstract}
\noindent
We discuss a class of models of neutrino masses and mixings with inverse hierarchy based on a broken $U(1)_F$ flavour symmetry with charge $L_e-L_\mu-L_\tau$ for lepton doublets and arbitrary right-handed charges. The symmetry breaking sector receives separate contributions from flavon vev breaking terms and from soft mass breaking in the right handed Majorana sector. The model is able to reproduce in a natural way all observed features of the charged lepton mass spectrum and of neutrino masses and mixings (even with arbitrarily small $\theta_{13}$), with the exception of a moderate fine tuning which is needed to accomodate the observed small value of $r =  \Delta m^2_{sol} /\Delta m^2_{atm}$.

\end{abstract}
\end{titlepage}
\setcounter{footnote}{0}
\vskip2truecm
%%%%%%%%%%%%%%%%%%%%%%%% 1.  INTRODUCTION   %%%%%%%%%%%%%%%%%%%%%%%%%%%%%%
%
\section{Introduction}
As well known \cite{Altarelli:2004za}, a simple dynamical approach to construct  models of neutrino masses and mixings with inverse hierarchy is to start in first approximation from a $U(1)_F$ flavour symmetry \cite{Froggatt:1978nt} with charge $L_e-L_\mu-L_\tau$ \cite{Petcov:1982ya}, where $L_i$ are the separate lepton numbers with flavour $i$. If this charge is assigned to the 3 families of $SU(2)$ lepton doublets $l$,  the neutrino mass matrix $m_\nu \sim l^Tl$, in the limit of exact symmetry, is of the form
\beq m_\nu =m
\left(
\begin{array}{ccc}
0& 1& x\\ 1& 0&0\\ x& 0&0
\end{array}
\right)~~~~~~~~~~~~~,
\label{invh1}
\eeq
The eigenvalues are $(m_1,m_2,m_3) \sim (m,-m,0)$, which indeed correspond to an inverse hierarchy spectrum with $m^2 \sim \Delta m^2_{atm}$. The opposite sign of $m_1$ and $m_2$ makes the spectrum sufficiently stable under renormalisation group corrections \cite{Chankowski:2001mx} which can then be neglected for our purposes. In the (generally unrealistic) assumption of diagonal charged leptons, the mixing angles would be given by $\theta_{12} = \pi/4$ (i.e.  maximal solar angle), $\tan{\theta_{23}}= x$ (i.e. generically large atmospheric angle) and $\theta_{13} = 0$. As the observed mixing angles show the opposite pattern, with a definitely non maximal
solar angle and an atmospheric angle compatible with being maximal \cite{fogli}, a substantial correction to the mixing angles is required from this starting approximation. The physical value of $r =  \Delta m^2_{sol} /\Delta m^2_{atm} \sim 1/30$ \cite{fogli}, which is zero in the symmetric limit, must also be obtained from the corrective terms. A main problem is that simple forms of $U(1)_F$  symmetry breaking tend to produce too small a deviation of the solar angle from its maximal value \cite{Grimus:2000kv,He:2002rv,Barbieri:1998mq}. However, the correction to bring the solar angle down from its maximal value can be obtained as an
effect of the charged lepton matrix diagonalization \cite{Petcov:2004rk}. 
This possibility, studied in detail in refs. \cite{noilast,frampton,Romanino:2004ww}, in abstract terms is not excluded but is strongly constrained by the observed smallness of $\theta_{13}$, as the amount of deviation from a maximal solar angle is of order $\theta_{13}$. This mechanism can only work if the value  of $\theta_{13}$ is very close to its present upper bound. Within the $U(1)_F$ framework the question of which symmetry breaking mechanism is capable of leading to a suitable charged lepton mass matrix for this purpose is left completely unanswered.

Recently Grimus and Lavoura  (GL) \cite{GriLa} suggested that, in a see-saw realization of the model,  a large breaking of $U(1)_F$ could be present in $M_{RR}$, the right-handed neutrino Majorana matrix.  A crucial point is that a large symmetry breaking in $M_{RR}$ cannot propagate via radiative corrections to other sectors of the lagrangian given that $\nu_R$ has no gauge interactions. One could possibly reproduce the observed values of the solar and atmospheric mixing angles together with a small value of $\theta_{13}$ (which could  even be zero). We present here a detailed discussion of models based on this idea. With respect to GL there are important additions and differences: we have three right-handed neutrinos, we explicitly discuss the charged lepton sector (in particular, we aim at reproducing the mass hierarchy in a natural way) and we complete the model with additional structure at the non leading level.  The large breaking of $U(1)_F$ from $M_{RR}$ produces the required large correction to fix $\theta_{12}$ with $\theta_{13}$ still vanishing at this level.  An additional, more conventional,  form of $U(1)_F$ symmetry breaking in terms of  vev's of a flavon field produces a pattern of non leading corrections. These terms provide the correct hierarchy for the charged lepton masses and, through the effect of the charged lepton diagonalization, shift the $\theta_{13}$ value away from zero by a small amount, of the order of a  power of charged lepton mass ratios. In this way a considerable freedom in the possible range of  $\theta_{13}$ is made compatible with the observed values of the solar and atmospheric mixing angles. However, a moderate fine tuning is necessary to reproduce the observed smallness of $r$ that normally should be $\sim o(1)$ in this kind of models.

\section{Models with Broken $L_e - L_\mu - L_\tau$ }

In the leptonic sector, we adopt the following classification under $U(1)_F$: 
\bea
l_i\sim L_e-L_\mu-L_\tau \sim (1,-1,-1)\nonumber \\
l_{Ri} \sim (Q_e, Q_\mu, -1)\nonumber\\
\nu_{Ri}\sim (-Q_R,Q_R,0)\label{qfl}
\eea
The Higgs doublet $H$ is taken to carry $Q_H=0$. In presence of only one Higgs doublet this does not imply a loss of generality. Note that, for right handed charged leptons, the $U(1)_F$ charge assignment does not coincide with $L_e-L_\mu-L_\tau$ (the latter would lead to no structure for charged lepton masses).  The $U(1)_F$ symmetry is broken by the vev of a complex field $\theta$ of charge $Q_\theta=1$ (and $\theta^\dagger$ has, of course, charge $-1$, so that one has the same suppression for either sign of the charge mismatch). The suppression factor for a unit charge mismatch is $\lambda=
<\theta>/\Lambda$, with $\Lambda$ being the cut-off. The resulting charged lepton mass matrix is of the form (with coefficients of o(1) at each entry always understood as in all $U(1)_F$ models):
\beq
m^l\sim \bar l l_R \sim m_\tau \left(
\begin{array}{ccc}
\lambda^{|-1+Q_e|}& \lambda^{|-1+Q_\mu|}& \lambda^2\\
\lambda^{|1+Q_e|}& \lambda^{|1+Q_\mu|}& 1\\
\lambda^{|1+Q_e|}& \lambda^{|1+Q_\mu|}& 1\\
\end{array}
\right)~~~~\sim~~ m_\tau \left(
\begin{array}{ccc}
\xi ^\prime&\xi \epsilon&\epsilon\\
\xi ^\prime \epsilon& \xi&1\\
\xi ^\prime \epsilon& \xi&1
\end{array}
\right)~~.\label{melle}
\eeq
where we have set
\beq
\epsilon \sim \lambda^2,~~~~\xi \sim  \lambda^{|1+Q_\mu|}\sim \frac{m_\mu}{m_\tau}\sim 6~10^{-2},~~~~\xi^\prime \sim  \lambda^{|-1+Q_e|}\sim \frac{m_e}{m_\tau}\sim 3~10^{-4}.
\label{def}
\eeq
Note that the charged lepton mass ratios are almost insensitive to the renormalization group scale from the electroweak up to the GUT energy \cite{Fusaoka:1998vc}. In eq. (\ref{melle}) we have taken $Q_\tau=-1$ in order to make $m_\tau \sim o(1)$. The values of $\lambda$, $Q_e$ and $Q_\mu$ are to be chosen in such a way as to reproduce the correct pattern of masses. In particular the signs of $Q_e$ and $Q_{\mu}$ are fixed in order to obtain a charged lepton mass matrix with the structure shown in eq. (\ref{mellediag}). Also, we will see in the following that $\theta_{13}$ will end up being of order $\lambda$, so that we take $\lambda < ~ 0.25$. Then a set of possible values is shown in Table 1.
\begin{table}[htdp]
\begin{center}
\begin{tabular}{|c|c|c|c|}
\hline 
$\lambda$&
$Q_{e}$&
$Q_{\mu}$&
$Q_{\tau}$\tabularnewline
\hline
\hline 
$0.25\sim\xi^{\frac{1}{2}}$&
7&
-3&
-1\tabularnewline
\hline 
$0.15\sim\xi^{\frac{2}{3}}$&
$\frac{11}{2}$&
$-\frac{5}{2}$&
-1\tabularnewline
\hline 
$0.06\sim\xi$&
4&
-2&
-1\tabularnewline
\hline 
$4\times10^{-3}\sim\xi^{2}$&
$\frac{5}{2}$&
$-\frac{3}{2}$&
-1\tabularnewline
\hline 
$2\times10^{-4}\sim\xi^{3}$&
2&
$-\frac{4}{3}$&
-1\tabularnewline
\hline
\end{tabular}
\end{center}
\label{default}
\caption{A set of charges and corresponding values of $\lambda$ that can accomodate the observed charged lepton masses. In the following, we will argue that, in this class of models, we expect $\theta_{13} \leq o(\lambda)$, so that we have restricted us to values of $\lambda$ below the upper bound for $\theta_{13}$.}

\end{table}

The charged lepton mass matrix, eq(\ref{melle}), can be related to its diagonal form by multiplication on the left by:
\beq
U_l=R_{23}(\theta_l)R_{13}(\epsilon)R_{12}(\epsilon)\label{Ul}
\eeq
where $R_{ij}(\theta)$ is a rotation in the $ij$ plane of angle $\theta$, $\theta_l$ is an angle of $o(1)$ and by $\epsilon$ we indicate an angle of order $\epsilon \sim \lambda^2$ \cite{Barbieri:1998mq}. The relation is fixed by:
\beq
 \left(
\begin{array}{ccc}
\xi ^\prime&\xi \epsilon&\epsilon\\
\xi ^\prime \epsilon& \xi&1\\
\xi ^\prime \epsilon& \xi&1
\end{array}
\right)~=~U_l~\left(
\begin{array}{ccc}
\xi ^\prime&0&0\\
0& \xi&0\\
0&0&1
\end{array}
\right).\label{mellediag}
\eeq
The matrix $U_l$ contributes to the neutrino mixing matrix via the relation $U_{PNMS}=U_l^\dagger U_\nu$. The resulting effect is a shift of  $\theta_{23}$ by a large angle proportional to $\theta_l$, while $\theta_{12}$ and $\theta_{13}$ are corrected by small terms of o($\epsilon \sim \lambda^2$). We shall see that larger contributions of order $\lambda$ to $\theta_{13}$ may arise from the neutrino sector in the present model. We clearly see that, in the framework of $U(1)_F$ broken by vev's, $\theta_{13}$ is normally forced to be small by the observed values of the charged lepton mass ratios. This is because $\theta_{13} \sim o(\lambda^2)$ and $\lambda$ must be small in order for $\xi$ and $\xi'$ to reproduce the observed mass ratios.

We now consider the neutrino mass sector. The value  of the charge $Q_R$ in eq. (\ref{qfl}) must be specified. Only for an absolute value $|Q_R|=1$ we obtain that some Dirac neutrino mass matrix elements are unsuppressed, and the sign choice actually leads to equivalent models for all practical purposes (after some parameter reshuffling and neglecting second order effects in the $U(1)_F$ breaking). Thus we take $Q_R=1$ in the following. The neutrino Dirac matrix is then of the form:
\beq
m^D_\nu \sim \bar \nu_R~l  \sim m \left(
\begin{array}{ccc}
y_{11}\lambda^2& a&b\\
d&y_{22}\lambda^2& y_{23}\lambda^2\\
y_{31}\lambda&y_{32}\lambda& y_{33}\lambda\\
\end{array}
\right)~~.\label{mDnu}
\eeq
Here we have esplicitly indicated the coefficients $y_{ij}$, $a$, $b$ and $d$, all of o(1).
Similarly for the neutrino right-handed Majorana matrix we should write:
\beq
m^{-1}_{RR}  \sim \frac{1}{M} \left(
\begin{array}{ccc}
x_{11}\lambda^2&W&x_{13}\lambda\\
W&x_{22}\lambda^2& x_{23}\lambda\\
x_{13}\lambda&x_{23}\lambda& Z\\
\end{array}
\right)~~.\label{mRRla}
\eeq
Note that, in this case, the orders of magnitude of the various entries would be the same in $m^{-1}_{RR}$ and $m_{RR}$.
But, as explained in the introduction, we assume, following GL, that in $m_{RR}$ a different source of $U(1)_F$ breaking could be operating in addition to that induced by the flavon vev's, in the form of explicit soft mass terms. We thus write down a generic form for $m^{-1}_{RR}$:
\beq
m^{-1}_{RR}  \sim \frac{1}{M} \left(
\begin{array}{ccc}
A&W&B\\
W&C& D\\
B&D& Z\\
\end{array}
\right)~~.\label{mRR}
\eeq
In the notation adopted $W$ and $Z$ are the terms that are already present in the symmetric limit, while $A$, $B$, $C$ and $D$ are symmetry breaking terms. While we keep in mind, for use in the following, that in the case of $U(1)_F$ broken by flavon vev's, $|A|$,$|C|\sim \lambda^2$ and $|B|$,$|D|\sim \lambda$, we make no statements about the orders of magnitude of each entry at this stage.
From the see-saw formula we obtain:
\bea
m_\nu&=&m^{DT}_\nu m^{-1}_{RR}m^D_\nu   \sim m_{\nu 0} + \lambda m_{\nu 1}+\dots \sim \cr  \cr
&\sim& \frac{m^2}{M} \left(
\begin{array}{ccc}
d^2C&adW&bdW\\
adW&a^2A&abA\\
bdW&abA&b^2A\\
\end{array}
\right)~+ \cr\cr
&+&~\lambda~  \frac{m^2}{M} \left(
\begin{array}{ccc}
2y_{31}dD&y_{31}aB+y_{32}dD&y_{31}bB+y_{33}dD\\
y_{31}aB+y_{32}dD&2y_{32}aB&y_{32}bB+y_{33}aB\\
y_{31}bB+y_{33}dD&y_{32}bB+y_{33}aB&2y_{33}bB\\
\end{array}
\right)~+~o(\lambda^2)~~~.\label{mnu}
\eea
We note that $Z$ does not appear at all to $o(\lambda)$, while $A$, $C$ and $W$ only appear in  $m_{\nu 0}$ and  $B$ and $D$ only appear in  $m_{\nu 1}$. Note that, in the case of $U(1)_F$ broken by flavon vev's, $|A|$,$|C|\sim \lambda^2$ and $|B|$,$|D|\sim \lambda$, so that all entries in $m_{\nu 0} + \lambda m_{\nu 1}$ are either of $o(1)$ (the $W$ terms) or of $o(\lambda^2)$. In this case the pattern for the neutrino mass matrix would be exactly the same even for no right-handed neutrinos and no see-saw, with $m_\nu \sim l^Tl$ obtained directly from the $L_e-L_\mu-L_\tau$ charges of the lepton doublets $l$. Instead, if the dominant breaking of $U(1)_F$ is through mass terms in $m_{RR}^{-1}$, the see-saw generation of masses is clearly essential and the resulting breaking can be larger, for example by terms of $o(1)$ in $m_{\nu 0}$ (the $A$ and $C$ terms), and by terms of $o(\lambda)$ in $\lambda m_{\nu 1}$ (the $B$ and $D$ terms).

We can analyse the results at various levels of symmetry breaking. For exact $U(1)_F$ we can set $\lambda=0$ and keep only the terms in $W$, $a$, $b$ and $d$. At this level we get the well known predictions of exact $L_e-L_\mu-L_\tau$:
\beq
m_1=~-~m_2;~~~~m_3=0;~~~~\theta_{13}=0;~~~~\theta_{12}=\frac{\pi}{4};~~~~ \tan{\theta_{23}}=\frac{b}{a}.\label{00}
\eeq
At the next level, we only take $m_{\nu0}$ in eq.(\ref{mnu}), i.e. we drop all $\lambda$ terms and only keep the symmetry breaking terms from  $A$ and $C$. Note that in general there are three phases left in the physical light neutrino mass matrix. However, since for $m_{\nu0}$ both $m_3$ and $\theta_{13}$ vanish, only one single phase is left. We can choose, for example, to take $\bar W$ and $\bar A - \bar C$ real and write $\bar A +\bar C= - |\bar A +\bar C| e^{i\delta}$.
We then obtain:
\bea
m_3=0 ;& m_1+m_2=\bar C + \bar A;& ~m_1 - m_2 =\sqrt{(\bar A-\bar C)^2+\bar W^2};\cr
\theta_{13}=0;& ~\tan{\theta_{23}}=\left| \frac{b}{a} \right|;& ~\tan^2{2\theta_{12}} \sim \frac{\bar W^2}{(\bar A-\bar C)^2} \label{01}
\eea
where we have set:
\beq
\bar A = \frac{m^2}{M}~A~(a^2+b^2);~~~~\bar C = \frac{m^2}{M}~C~d^2;~~~~\bar W^2 = \frac{m^2}{M} 4W^2 (a^2+b^2) d^2.\label{defs}
\eeq

We stress that, at this level, all results are completely equivalent to those obtained in the model of GL where only two right-handed neutrinos are assumed. In fact, the presence of a third $\nu_R$ does not introduce additional parameters as only $A$, $C$ and $W$ from the 1-2 submatrix of $m^{-1}_{RR}$ appear in  $m_{\nu0}$.
At this stage we see that in order to obtain a realistic value for $\tan^2{2\theta_{12}}$ we need the symmetry breaking parameter $|\bar A-\bar C|$ to be not too small with respect to the symmetric term $\bar W$. Experimentally, at 2-$\sigma$, $\sin^2{\theta_{12}}\sim 0.314+0.06-0.05$ \cite{fogli} so that $\tan^2{2\theta_{12}} = t^2 \sim 6$, and we obtain:
\beq
\frac{|\bar A-\bar C|}{|\bar W|} \sim 0.40.\label{cond1}
\eeq
On the other hand, $r$ is given by:
\beq
r =( |m_2|^2- |m_1|^2)/  |m_2|^2= \frac{4|\bar A+\bar C| \cos{\delta}\sqrt{(\bar A-\bar C)^2+\bar W^2}}{|\bar A+\bar C+\sqrt{(\bar A-\bar C)^2+\bar W^2}|^2}\sim \frac{4\eta \cos{\delta} \sqrt{1+t^2}}{|-\eta e^{i\delta}+\sqrt{1+t^2}|^2}.\label{r}
\eeq
Note that we take $\cos{\delta}>0$ in order that $|m_2|^2- |m_1|^2 >0$ in agreement with the usual definition of $m_{1,2}$. The numerical values for $\eta$ and $\cos{\delta}$ are constrained by the observed value of $r$, $r \sim 1/30$ \cite{fogli}. For example, if $\eta$ is small one finds approximately:
\beq
\eta \cos{\delta}= |\frac{\bar A+\bar C}{\bar A-\bar C}|\cos{\delta}\sim 0.02.\label{eta}
\eeq
We have discarded  the possibility of very large $\eta$ because $m_1+m_2=\bar A + \bar C$ is zero in the exact symmetry limit. We see that a moderate amount of fine tuning is needed to fix the observed value of $r$ which would naturally be expected to be of $\sim o(1)$ in this model. 

We now consider the effects at order $\lambda$ induced by the term $\lambda m_{\nu 1}$ in eq.(\ref{mnu}). First of all, in general all quantities which were vanishing for $\lambda=0$ are now different from zero. In particular, $\theta_{13}$ receives a contribution at order $\lambda$ given by:
\beq
|\tan{\theta_{13}}|\sim |\lambda \left[\frac{B}{W}-\frac{A}{W}\frac{D}{W}\right] \frac{y_{33}a-y_{32}b}{d\sqrt{a^2+b^2}}|.\label{la13}
\eeq
As we have already mentioned, we expect $B$ and $D$ in the range $o(\lambda) \leq |B|,~|D| \leq o(1)$, the lower limit corresponding to the flavon vev breaking of $U(1)_F$ and the upper limit to the size of the soft breaking terms $A$ and $C$. Thus $\tan{\theta_{13}}$ acquires a non vanishing value $o(\lambda^2) \leq  |\tan{\theta_{13}}| \leq o(\lambda)$. We recall that a contribution of $o(\lambda^2)$ is in any case expected from the diagonalization of the charged lepton matrix. Going back to Table 1 we see that all measurable values for $\theta_{13}$ can be obtained with a suitable choice of the $U(1)_F$ charges as well as values below the experimentally accessible range. Thus, in this class of models, there is no difficulty in accomodating a large deviation of the solar angle $\theta_{12}$ from the maximal value together with a size of $\theta_{13}$ much below the present experimental bound. Another quantity which vanishes for $\lambda=0$ is $m_3$. It is easy to see that $m_3$ remains zero at $o(\lambda)$ and only receives a contribution at next order: $m_3 \sim o(\lambda^2)$. In fact, the neutrino Dirac mass matrix $m_\nu^D$ in eq.(\ref{mDnu}) has a determinant of $o(\lambda)$ so that $m_\nu$ obtained from the see-saw formula in eq.(\ref{mnu}) has a determinant of $o(\lambda^2)$. Thus the corrected mass spectrum is still of the inverse hierarchy type. The quantity $r=\Delta m^2_{12}/\Delta m^2_{23}$ is corrected by terms of $o(\lambda)f_r(B,D)$, where $f_r(B,D)\sim o(\lambda) - o(1)$. Given that the observed value of $r$ is small, $r \sim 1/30$, if $f_r(B,D) \sim o(1)$ then the largest values of $\lambda$ shown in Table 1 would be disfavoured, and the case $ \theta_{13} \sim r$ or smaller would be indicated, barring cancellations or additional fine tuning. All other quantities like $\theta_{12}$, $\theta_{23}$ etc receive corrections of $o(\lambda)f(B,D)$, which however are not important as the leading terms are $o(1)$.

\section{Conclusion}

In this note we have reconsidered models of neutrino masses with inverse hierarchy based on a broken $U(1)_F$ symmetry with charge $Q_l=L_e-L_\mu-L_\tau$. These models can reproduce many of the observed features of the charged lepton spectrum and of the neutrino mixing angles, but a completely natural model is difficult to realize. The problem is that, in the limit of unbroken $U(1)_F$, $1-\tan{\theta_{12}}$, $r$ and $\theta_{13}$ are all zero. This is also true for the charged lepton mass ratios $m_e/m_\tau$ and  $m_\mu/m_\tau$ for our choice of right-handed charges (motivated by the observed smallness of these ratios). Experimentally, however, $1-\tan{\theta_{12}}$ turns out to be large. The difficulty is to have a large symmetry breaking correction for this particular observable while keeping all other deviations from the symmetric limit small. In fact, if the symmetry breaking occurs through a number of flavon vev's, the symmetry breaking order parameter $\lambda$ must be small to fit the charged lepton mass ratios as well as $r$ and $\theta_{13}$. Then $1-\tan{\theta_{12}}$ in general would also be small. The idea of disentangling $\theta_{12}$ from the maximal value as an effect of the charged lepton diagonalization involves some amount of stretching, because it needs $\theta_{13}$ to be very near its present upper bound, and, also, the question of why they turn out to be rather large is usually not addressed. The version of the model, studied in detail here, where, to the usual breaking of $U(1)_F$ by vev's, one adds a large source of symmetry breaking in the $m_{RR}$ mass sector is remarkable, because it allows a natural separation of $1-\tan{\theta_{12}}$, which indeed becomes large, from the charged lepton mass ratios and $\theta_{13}$, which are naturally predicted to be small. The only remaining imperfection is that, in this case,  $1-\tan{\theta_{12}}$ and $r$ should be of the same order, in particular both of $o(1)$ while they differ by a factor of about 10. This moderate fine tuning is the price to pay for an otherwise natural model with a simple structure. Finally, in this model, for $m_{ee}$, the parameter measured in neutrinoless double beta decay, one obtains at 2$\sigma$ the range $0.012\leq |m_{ee}| \leq 0.054$, that could be reached by the next generation of experiments.


\vfill


\newpage
\begin{thebibliography}{99}

\bibitem{Altarelli:2004za}
For recent reviews see for example G. Altarelli and F. Feruglio,
\newblock New J. Phys. 6 (2004) 106, \hepph{0405048};
%%CITATION = HEP-PH 0405048;%% 
S.~F.~King,
  %``Neutrino mass models,''
  Rept.\ Prog.\ Phys.\  {\bf 67} (2004) 107
  \hepph{0310204};
  %%CITATION = HEP-PH 0310204;%%
  W.~Grimus,
  %``Neutrino physics: Theory,''
  Lect.\ Notes Phys.\  {\bf 629} (2004) 169
  \hepph{0307149};
  %%CITATION = HEP-PH 0307149;%%
  I. Dorsner and S.M. Barr,
\newblock Nucl. Phys. B617 (2001) 493, \hepph{0108168}.
%%CITATION = HEP-PH 0108168;%% 
    

\bibitem{Froggatt:1978nt}
C.D. Froggatt and H.B. Nielsen,
\newblock Nucl. Phys. B147 (1979) 277.
%%CITATION = NUPHA,B147,277;%%

\bibitem{Petcov:1982ya}
  S.~T.~Petcov,
  %``On Pseudodirac Neutrinos, Neutrino Oscillations And Neutrinoless Double
  %Beta Decay,''
  Phys.\ Lett.\ B {\bf 110} (1982) 245.
  %%CITATION = PHLTA,B110,245;%%

\bibitem{Chankowski:2001mx}
  S.~Antusch, J.~Kersten, M.~Lindner and M.~Ratz,
  %``Neutrino mass matrix running for non-degenerate see-saw scales,''
  Phys.\ Lett.\ B {\bf 538} (2002) 87
  \hepph{0203233};
  %%CITATION = HEP-PH 0203233;%% 
  M.~Frigerio and A.~Y.~Smirnov,
  %``Radiative corrections to neutrino mass matrix in the standard model and
  %beyond,''
  JHEP {\bf 0302} (2003) 004
  \hepph{0212263};
  %%CITATION = HEP-PH 0212263;%% 
  S.~Antusch, J.~Kersten, M.~Lindner and M.~Ratz,
  %``Running neutrino masses, mixings and CP phases: Analytical results and
  %phenomenological consequences,''
  Nucl.\ Phys.\ B {\bf 674} (2003) 401
  \hepph{0305273};
  %%CITATION = HEP-PH 0305273;%% 
  J.~A.~Casas, J.~R.~Espinosa and I.~Navarro,
  %``Large mixing angles for neutrinos from infrared fixed points,''
  JHEP {\bf 0309} (2003) 048
  \hepph{0306243};
  %%CITATION = HEP-PH 0306243;%%  
  A.~Broncano, M.~B.~Gavela and E.~Jenkins,
  %``Renormalization of lepton mixing for Majorana neutrinos,''
  Nucl.\ Phys.\ B {\bf 705} (2005) 269
  \hepph{0406019};
  %%CITATION = HEP-PH 0406019;%%  
  C.~Hagedorn, J.~Kersten and M.~Lindner,
  %``Stability of texture zeros under radiative corrections in see-saw
  %models,''
  Phys.\ Lett.\ B {\bf 597} (2004) 63
  \hepph{0406103}.
  %%CITATION = HEP-PH 0406103;%%
  
  For a review on older works see P.H. Chankowski and S. Pokorski,
\newblock Int. J. Mod. Phys. A17 (2002) 575, \hepph{0110249}.
%%CITATION = HEP-PH 0110249;%%

\bibitem{fogli}
G.L. Fogli et~al.,
\newblock (2005), \hepph{0506083}.
%%CITATION = HEP-PH 0506083;%%

\bibitem{Grimus:2000kv}
W. Grimus and L. Lavoura,
\newblock Phys. Rev. D62 (2000) 093012, \hepph{0007011};
%%CITATION = HEP-PH 0007011;%% 
L. Lavoura and W. Grimus,
\newblock JHEP 09 (2000) 007, \hepph{0008020};
%%CITATION = HEP-PH 0008020;%% 
Q. Shafi and Z. Tavartkiladze,
\newblock Phys. Lett. B482 (2000) 145, \hepph{0002150};
%%CITATION = HEP-PH 0002150;%% 
S.F. King and N.N. Singh,
\newblock Nucl. Phys. B596 (2001) 81, \hepph{0007243};
%%CITATION = HEP-PH 0007243;%%
K.S. Babu and R.N. Mohapatra,
\newblock Phys. Lett. B532 (2002) 77, \hepph{0201176};
%%CITATION = HEP-PH 0201176;%%
  P.~Langacker and B.~D.~Nelson,
  %``String-inspired triplet see-saw from diagonal embedding of SU(2)L in SU(2)A
  %x SU(2)B,''
  Phys.\ Rev.\ D {\bf 72} (2005) 053013
  \hepph{0507063}.
  %%CITATION = HEP-PH 0507063;%%

\bibitem{He:2002rv}
H.J. He, D.A. Dicus and J.N. Ng,
\newblock Phys. Lett. B536 (2002) 83, \hepph{0203237}.
%%CITATION = HEP-PH 0203237;%%

\bibitem{Barbieri:1998mq}
R. Barbieri et~al.,
\newblock JHEP 12 (1998) 017, \hepph{9807235}.
%%CITATION = HEP-PH 9807235;%%

\bibitem{Petcov:2004rk}
S.T. Petcov and W. Rodejohann,
\newblock Phys. Rev. D71 (2005) 073002, \hepph{0409135};
%%CITATION = HEP-PH 0409135;%%
F. Feruglio,
\newblock Nucl. Phys. Proc. Suppl. 143 (2005) 184, \hepph{0410131}.
%%CITATION = HEP-PH 0410131;%%

\bibitem{noilast}
G. Altarelli, F. Feruglio and I. Masina,
\newblock Nucl. Phys. B689 (2004) 157, \hepph{0402155}.
%%CITATION = HEP-PH 0402155;%%

\bibitem{frampton}
P.H. Frampton, S.T. Petcov and W. Rodejohann,

\newblock Nucl. Phys. B687 (2004) 31, \hepph{0401206}.
%%CITATION = HEP-PH 0401206;%%

\bibitem{Romanino:2004ww}
A. Romanino,
\newblock Phys. Rev. D70 (2004) 013003, \hepph{0402258}.
%%CITATION = HEP-PH 0402258;%%

\bibitem{GriLa}
W. Grimus and L. Lavoura,
\newblock (2004), \hepph{0410279}.
%%CITATION = HEP-PH 0410279;%%

\bibitem{Fusaoka:1998vc}
  H.~Fusaoka and Y.~Koide,
  %``Updated estimate of running quark masses,''
  Phys.\ Rev.\ D {\bf 57} (1998) 3986
  \hepph{9712201}.
  %%CITATION = HEP-PH 9712201;%%

 
 \end{thebibliography}
\end{document}